%% file: cvpr.tex
\documentclass[final]{cvpr}
\pdfoutput=1
\usepackage{times}
\usepackage{epsfig}
\usepackage{graphicx}
\usepackage{amsmath}
\usepackage{amssymb}
\usepackage{mathtools}

\usepackage{cuted}
\usepackage{capt-of}

\DeclarePairedDelimiter{\norm}{\lVert}{\rVert} 

\usepackage[pagebackref=true,breaklinks=true,colorlinks,bookmarks=false]{hyperref}

\newcommand\blfootnote[1]{%
  \begingroup
  \renewcommand\thefootnote{}\footnote{#1}%
  \addtocounter{footnote}{-1}%
  \endgroup
}

\def\cvprPaperID{3332} 
\def\confYear{CVPR 2021}

\begin{document}
\title{StyleUV: Diverse and High-quality UV Map Generative Model}

\author{

Myunggi Lee\thanks{These authors contributed equally.}${^{\,\,\,1,3}}$
\\ \vspace{-12pt}

\and
\hspace{-2pt}Wonwoong Cho\footnotemark[1]${^{\,\,\,2,3}}$
\\ \vspace{-12pt}

\and
\hspace{-2pt}Moonheum Kim${^3}$
\\ \vspace{-12pt}

\and
\hspace{-2pt}David Inouye${^2}$
\\ \vspace{-12pt}

\and
\hspace{-2pt}Nojun Kwak${^1}$
\\ \vspace{-12pt}

\and
${^1}$ Seoul National University
\and
${^2}$ Purdue University
\and
${^3}$ \fontsize{11.5pt}{11.5pt}\selectfont
{NAVER WEBTOON}
}

\twocolumn[{%
\renewcommand\twocolumn[1][]{#1}%
\maketitle

\vspace*{-1cm}
\begin{center}
    \centering
    
    \includegraphics[width=\linewidth]{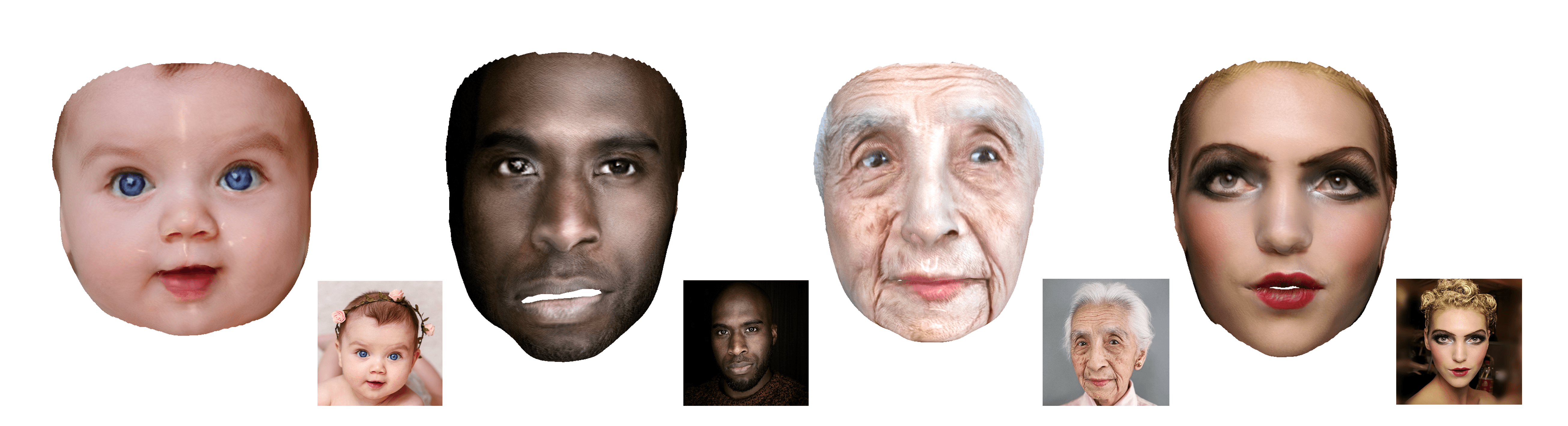}
    \captionof{figure}{Fitting results of our novel texture generative model. StyleUV can generate high-fidelity images and cover the diverse nature of human faces including but not limited to the faces of a baby, a black person, an elderly person, and a young woman wearing heavy makeup.}
    \label{fig:image_intro_compress}
\end{center}%
}]





\begin{abstract}
    Reconstructing 3D human faces in the wild with the 3D Morphable Model (3DMM) has become popular in recent years. While most prior work focuses on estimating more robust and accurate geometry, relatively little attention has been paid to improving the quality of the texture model. Meanwhile, with the advent of Generative Adversarial Networks (GANs), there has been great progress in reconstructing realistic 2D images. Recent work demonstrates that GANs trained with abundant high-quality UV maps can produce high-fidelity textures superior to those produced by existing methods. However, acquiring such high-quality UV maps is difficult because they are expensive to acquire, requiring laborious processes to refine. In this work, we present a novel UV map generative model that learns to generate diverse and realistic synthetic UV maps without requiring high-quality UV maps for training. Our proposed framework can be trained solely with in-the-wild images (i.e., UV maps are not required) by leveraging a combination of GANs and a differentiable renderer. Both quantitative and qualitative evaluations demonstrate that our proposed texture model produces more diverse and higher fidelity textures compared to existing methods.
\end{abstract}


\input{sections/Introduction}
\input{sections/RelatedWorks}
\input{sections/Methods}

\input{sections/Experiments}

\begin{figure*}[!t]
  \includegraphics[width=\linewidth/2]{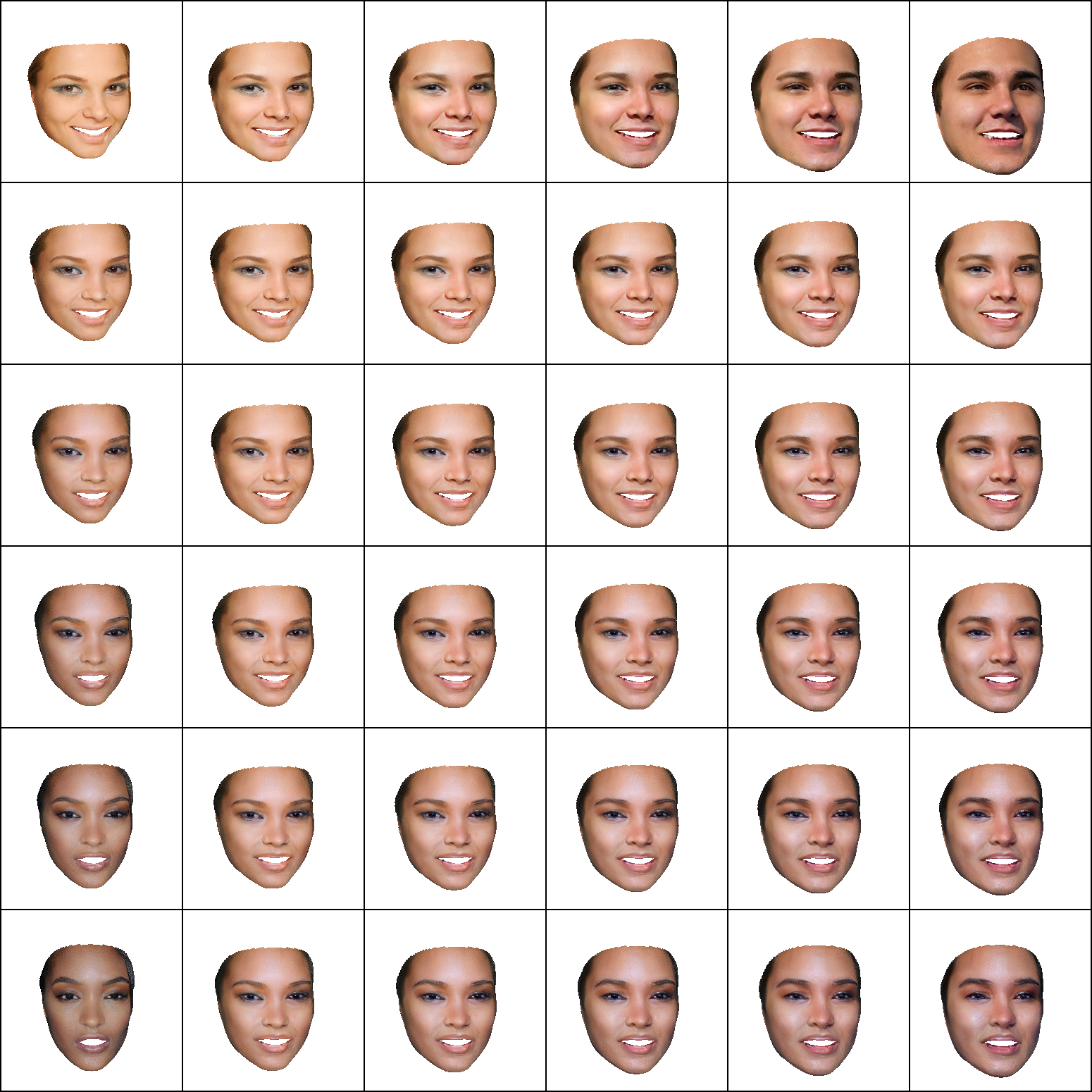}
  \includegraphics[width=\linewidth/2]{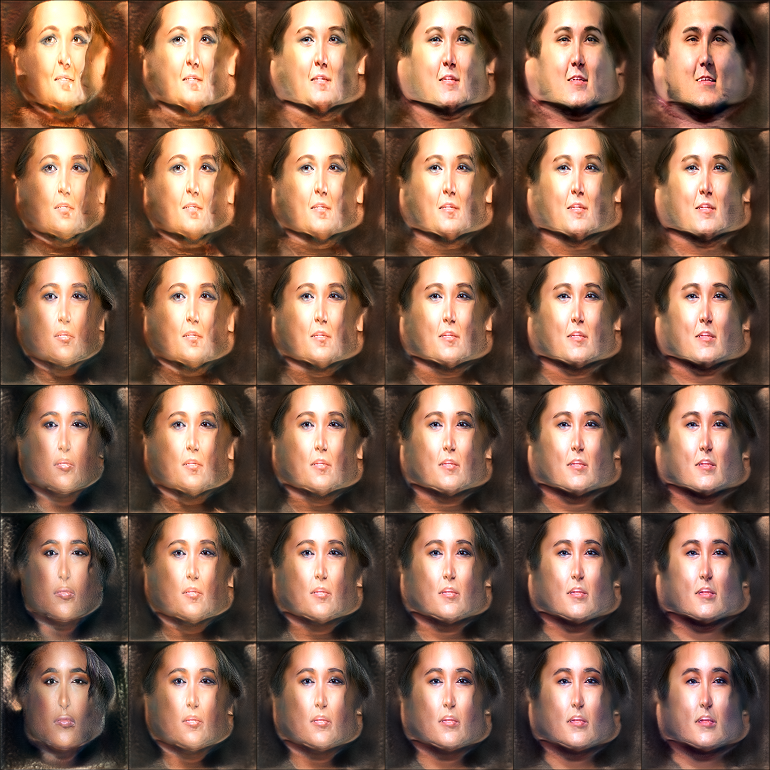}
  \caption{Feature interpolation. The left figure represents the projected images and the right figure shows the interpolated UV maps. The image in the left-top corner indicates a young woman wearing a makeup, the image in the right-top corner is a young man and the image in the left-bottom corner denotes a young black woman.}
\label{fig:sup_interpolation}
\end{figure*}

\begin{figure*}[!t]
  \includegraphics[width=\linewidth]{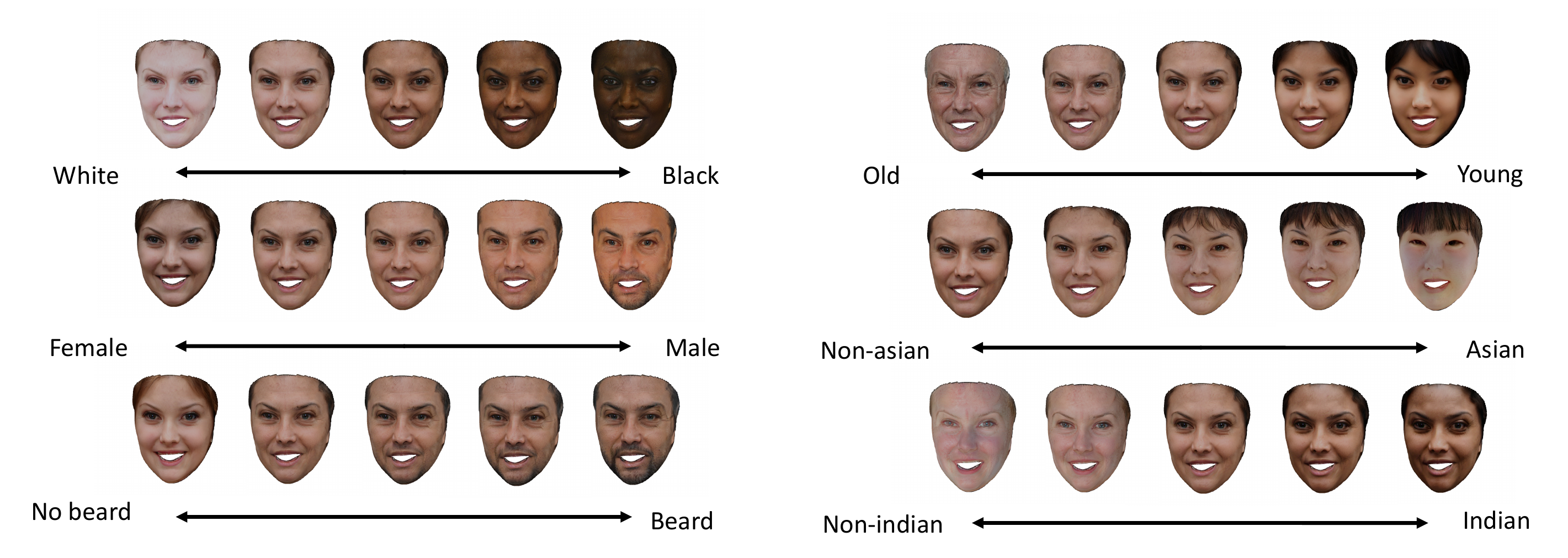}
  \caption{Semantic control over the various attributes.}
\label{fig:sup_attr_ctl}
\end{figure*}

\input{sections/Conclusion}

\section{Appendix}

Another strength of StyleUV lies in the disentangled nature of its latent space, i.e., the latent space of StyleUV is perceptually and semantically linear~\cite{Karras_2019_CVPR,Karras_2020_CVPR}. One of the benefits of the disentangled latent space is that a semantic manipulation is possible. 

\paragraph{Feature Interpolation}
We first qualitatively report the interpolation performance of StyleUV. Provided that any features that are not included in either end points appear in the midst of the interpolated points, it indicates the latent space is not semantically linear.

Fig.~\ref{fig:sup_interpolation} shows the interpolation results of StyleUV. To elaborate, the images in the top-left, top-right and bottom-left of both figures are fitted results. The right figure shows the results of bilinear interpolation within fitted UV maps, and the left figure represents the projected results of those UV maps. Note that 3D parameters such as ${p_i,p_e,p_c,p_l}$ are also interpolated when performing a projection. As represented in both figures, we can observe that the intermediate latent points represent well the semantics between the start and the end points. In case of the first row, for example, the makeup on the face is removed and the gender is changed from female to male, as going right side. This is because the leftmost image in the first row shows a young woman wearing a make-up while the rightmost image represents a young man. That is, this illustration demonstrates the latent space of StyleUV is semantically linear. 


\paragraph{Attribute Manipulation}
We also report the results of semantic manipulations of the UV map, based on InterFaceGAN~\cite{shen2020interfacegan}. Briefly, we first train an attribute classifier that takes an image and outputs a probability for the attribute. When training the binary classifier, we use labels from CelebA~\cite{liu2015faceattributes} and UTKFace~\cite{zhifei2017cvpr}. We then randomly sample 20,000 latent vectors and obtain corresponding UV maps by forwarding the latent vectors into the pretrained generator. After projecting the UV maps onto the image plane, we assign pseudo attribute labels for each latent vector by using the pretrained classifier on the UV maps. Employing the predefined datasets, we find a hyperplane that separates an attribute on the latent space\footnote{We use latent vector $w$ as the inputs for SVM since the latent space $\mathcal{W}$ behaves stable for the long term manipulation as reported in ~\cite{shen2020interfacegan}. $w$ denotes the output of Mapping Networks in StyleGAN v2.} by fitting a linear SVM to approximate labels for the latent vectors. Lastly, we perform a semantic manipulation of a latent vector by shifting it along a normal vector of the hyperplane.

As shown in Fig.~\ref{fig:sup_attr_ctl}, the results illustrate that the semantics of the UV maps can be controlled by navigating the latent space of StyleUV. This implies that the latent space of StyleUV is semantically disentangled, which has a potential to be usefully used in practice.

Overall, we believe the impressive results of StyleUV in both analyses can be reached due to our novel frameworks. That is, the well-defined latent space is achieved since the numerous training data, embracing the rich facial attributes of real world can be leveraged in StyleUV.

\clearpage

{\small
\bibliographystyle{ieee_fullname}
\bibliography{egbib}
}

\end{document}

%% file: sections/Introduction.tex
\section{Introduction}\label{sec:introduction}

Reconstruction of 3D face from a single RGB image has gained a plenty of attention in computer vision and graphics, due to its usefulness in diverse applications, such as
face recognition~\cite{Face-recognition-based,deepface}, face reenactment~\cite{kim2018deep} and forensics~\cite{schneider2019photo}. Notwithstanding its remarkable achievements in the last decades, a seamless texture model embracing both high-quality and diverse nature of the real world is beyond reach.

We posit that a couple of obstacles exist towards a breakthrough of the texture model.  
First, acquiring the massive number of 3D facial data containing abundant facial identities is limited because of its expensive price and arduous process.
Considering that human appearances have countless diversity, the finite number of available data hinders a 3D texture model from covering the real diversity.

\blfootnote{${^*}$These authors contributed equally.}

Second, existing models, such as a renowned linear 3D morphable model (3DMM) \cite{3DMM} and its variants~\cite{1613022,FaceWarehouse,gaussian_mixture_3dmm,LSFM}, have a clear limitation in representing a realistic and diverse face appearances. This is because they are modeled upon a small number of scan database, and the parametric subspace of the model occupies only a partial area of the entire texture space, from which point the diversity and the high-fidelity cannot be covered.

A slew of studies have explored to tackle the aforementioned problems by leveraging deep neural networks. Such non-linear methods are generally involved in the analysis-by-synthesis paradigm and designed to follow either a learning-based scheme~\cite{genova2018unsupervised,Tewari_2018_CVPR, tran2019towards, deng2019accurate, on-learning-3d-face-morphable-model-from-in-the-wild-images,tran2019towards} or an optimization-based scheme~\cite{Gecer_2019_CVPR}. With a combination of deep neural networks and the analysis-by-synthesis techniques, the reconstruction accuracy has been improved. However, the texture outputs still suffer from incorrect estimations that are blurry or non-photorealistic.

Arguably, generative adversarial networks~\cite{goodfellow2014generative} (GANs) have shown an unprecedentedly impressive performance in generating diverse and realistic images, which can be an exact remedy for the inaccurate texture model. In response, several previous studies~\cite{deng2018uv,nonlinear-3d-face-morphable-model,on-learning-3d-face-morphable-model-from-in-the-wild-images,lee2020uncertainty} based on the encoder-decoder structure have incorporated GANs' objective to their 3D reconstruction frameworks for enhancing the texture fidelity. Basically, a combination of cost functions based on self-supervision and the GANs' objective are used to train the networks. Those methods, however, are different from typical GANs~\cite{goodfellow2014generative,WGAN,wgan-gp}, that have a sampling space with a prior distribution and solely learn from the GANs' objectives.

The first research demonstrating a potential of GANs in texture generation is GANFIT~\cite{Gecer_2019_CVPR}. It proposed to use a GAN model to solve the analysis-by-synthesis problem. Briefly, a generator is pre-trained to output a realistic UV map. Next, it is employed as a texture model in its proposed reconstruction frameworks by finding an optimal latent vector of the fixed generator that best explains a given image in the analysis-by-synthesis manner. Although GANFIT has proven a promising aspect of GANs as a texture model for 3D face, the method relies on a real UV map dataset to obtain the pre-trained generator. However, the publicly available UV dataset~\cite{deng2018uv} has a critical weakness in either the limited number of identity or the lack of diversity, such as age (e.g., baby, young man, and elderly), race (e.g., white, black, Asian, etc.), and gender (e.g., female and male). 


To this end, we propose a novel UV map generative model that learns to generate diverse and realistic images in the analysis-by-synthesis manner. Briefly, our proposed networks learn to output a UV map by a combination of the state-of-the-art GAN architecture~\cite{Karras_2020_CVPR}, predefined 3D parameters~\cite{3DMM}, and geometric and photometric image formation functions~\cite{liu2019softras}.
The entire framework is fully differentiable, thus it is possible to train our generator with the GANs' objectives even after passing through the image formation functions.
Furthermore, since the generated image from the generator is a UV map, rather than an image itself, the background of the image (i.e., the part of the image that is not the face) is unnecessary and can be removed. 
In response to this systematic difference with the original GANs, we introduce a silhouette~\cite{liu2019softras} as a foreground mask, which brings a significant improvement of our model. 

Our technical novelties lie in three folds:
\begin{itemize}
\item{We propose a novel UV map generation framework that is able to generate photo-realistic and diverse UV maps. Instead of using the limited number of high-quality UV maps, we fully utilize high-quality images in the training phase with a differentiable renderer.}
\item{The proposed UV map generative model is a plug-and-play modular function, which is able to replace the linear texture model in 3DMM or other existing non-linear texture model.}
\item{We publicly open our novel texture model to accelerate the research field on texture generative model.}
\end{itemize}

%% file: sections/RelatedWorks.tex
\section{Related Work}
Reconstructing a 3D human face has been intensively studied for a decade since Blanz and Vetter introduced the 3D face model~\cite{3DMM} using principal component analysis (PCA) on 3D scans. Though unremitting exertion has been continued to improve 3DMM model fitting, less attention has been paid to enhancing the quality of texture model.

\vspace{1mm} 
\noindent \textbf{Linear Model}
Early works attempt to concentrate on regressing coefficients of 3DMM at firsthand~\cite{booth20173d, genova2018unsupervised,tuan2017regressing}. Booth et al.~\cite{booth20173d} proposed a feature-based texture model and Gauss-Newton iterative optimization to fit on an input image. Tran et al.~\cite{tuan2017regressing} regressed the shape and texture parameters directly from an input image by leveraging the power of the convolutional neural network (CNN). Genova et al.~\cite{genova2018unsupervised} tried to obtain the coefficients of 3DMM in an unsupervised manner utilizing the features from a facial recognition networks. More recently, Smith et al.~\cite{smith2020morphable} propose a 3D morphable albedo model which takes into account specular and diffuse maps for providing intrinsic high quality appearance regardless of the illumination condition. However, an inherent limitation is that the texture is generated within the boundary of 3DMM's texture space which resultingly brings a lack of high frequency details.

\vspace{1mm} 
\noindent \textbf{Non-linear Model}
To tackle the aforementioned problem, a major effort is under way to harness deep neural network for estimating 3D shape and texture reconstruction. Tewari et al.~\cite{Tewari_2018_CVPR} jointly learn parameters of 3DMM, skin reflectance and illumination in combination with a non-linear corrective model. Tran et al.~\cite{nonlinear-3d-face-morphable-model,on-learning-3d-face-morphable-model-from-in-the-wild-images} propose a novel framework to learn a non-linear 3DMM model from a large set of unconstrained images. They encode an input image into latent vectors which correspond to shape, texture and camera parameters respectively. Each of these latent vectors are further decoded to a 3D shape and a UV texture map through CNNs. Further improvement has been made by replacing the CNN with Graph Convolutional Network (GCN)~\cite{defferrard2016convolutional} for fine-detailed texture generation~\cite{lin2020towards, lee2020uncertainty}. One of the common features is that they formulate the problem as either regression methods or an auto-encoder architecture, which makes use of an image encoder and a differentiable renderer for self-supervision.

The recent work proposed by Gecer et al.~\cite{Gecer_2019_CVPR} revisits the optimization-based 3DMM fitting approach. They formulate a novel fitting strategy which is based on a GAN texture model pretrained on ground-truth UV dataset and a differentiable renderer. Although it makes a significant improvement in 3D texture recovery, it highly relies on a limited number of completed UV maps, which lacks diversity in representing in-the-wild images. Follow-up investigation has been pursued by Lee et al.~\cite{lee2020uncertainty} for further improvement. It uses a graph convolutional network combined with a UV map generation module used in~\cite{Gecer_2019_CVPR}. Most recent work by Lattas et al.~\cite{lattas2020avatarme} demonstrates a meticulously pipeline to reconstruct human faces with high frequency detail. They capture a large dataset of facial shape and reflectance and refine the state-of-the-art method using image-to-image translation framework. It makes a remarkable achievements in reconstructing render-ready 3D human faces.

\vspace{1mm} 
\noindent \textbf{High-quality Generative Model}
Since the advent of Generative Adversarial Network (GAN)~\cite{goodfellow2014generative} in 2014, there has been a significant progress in generating the photo-realistic images. Karras et al.~\cite{karras2017progressive} gathered a high-quality human face dataset CelebA-HQ and propose a progressive growing GAN that can generate realistic human faces at $1024^2$ scale. Brock et al.~\cite{brock2018large} proposed BIGGAN, which proves the large batch size brings performance improvements in high-resolution GAN training. More recently, Karras et al~\cite{Karras_2019_CVPR} collected more high-quality dataset FFHQ and proposed a style-based generative adversarial network (StyleGAN) which improves the quality of generated human face a stage further. Even though the computer vision community has witnesses the rapid development of 2D image generation, applying generative model to 3D vision tasks are still not fully explored.

%% file: sections/Methods.tex
\section{Backgrounds}
This section will describe the basic knowledge of 3DMM in subsection~\ref{subsection:3DMM}. Meanwhile, descriptions on UV map is provided in subsection~\ref{subsection:uv-representation}.


\subsection{3DMM}\label{subsection:3DMM}
3DMM~\cite{3DMM} is a widely-used linear statistical model that enables a 3D face with thousands of vertices to be estimated with only about a hundred of parameters. The linear statistical model comes from a multivariate normal distribution modeled upon hundreds of registered 3D face scanned data. The principal component analysis is then applied to the distribution, with which a formula for sampling out of the distribution can be represented as:
\begin{equation}
    S_{model}(p_s)=\Bar{S}+E_sp_s,\quad 
    T_{model}(p_t)=\Bar{T}+E_tp_t,
    \label{eq:pca}
\end{equation}
where $\{{\Bar{S},\Bar{T}\}\in \mathbb{R}^{3n}}$ are a mean shape and a mean texture, each of ${\{E_s, E_t\}\in \mathbb{R}^{3n\times k}}$ is a bunch of eigenvectors for the shape and the texture spanning the parametric space, and ${\{p_s,p_t\}\in \mathbb{R}^{k}}$ are the parameters to be optimized. ${n}$ is the number of vertices and ${k}$ is the number of basis vectors drawing a principal subspace where the 3D faces lie on. Note that a random variable for modeling the distribution of the shape model ${S_{model}}$ is set to the coordinate of vertices and it is set to RGB values when modeling the texture model ${T_{model}}$.

Regarding the shape model, it is further divided into an identity model and an expression model~\cite{FaceWarehouse}, so that the shape mean ${\Bar{S}}$ and the set of eigenvectors for the shape model ${E_s}$ are separated into ${\Bar{S}_i+\Bar{S}_e}$ and ${[E_i,E_e]\in\mathbb{R}^{3n\times (k_i+k_e)}}$ respectively, where ${k_i}$ and ${k_e}$ are the number of parameters determining the identity and the expression of a 3D face.

\vspace{1mm} 
\noindent \textbf{Limitations of linear texture model.} As pointed out by previous studies~\cite{nonlinear-3d-face-morphable-model,on-learning-3d-face-morphable-model-from-in-the-wild-images,Tewari_2018_CVPR,History}, the linear texture model has a clear weakness in representing a diverse and realistic texture because of its formulation~\cite{nonlinear-3d-face-morphable-model,Gecer_2019_CVPR} and the scarcity of available 3D facial data. Specifically, the limited number of 3D data hinders the texture model from learning the diverse nature of real texture of a human. Moreover, the low-dimensional subspace of the texture model also limits the expressive power of it. This is one of our strong motivations and will be compared to our proposed method in Section~\ref{section:experiments}. 

\subsection{UV Representation}\label{subsection:uv-representation}
UV map is a widely-used texture representation of 3D mesh. By using a predefined sampling grid, commonly generated by cylindrical unwarp, the RGB values of vertices can be unwrapped into a UV map. Suppose each vertex, $\{v_i\}_{i=1}^n$, in a 3D mesh has a texture color $t_i=[R_i, G_i, B_i]^T$ and a texture coordinate $c_i\in\mathbb{R}^{2}$ 
that assigns the texture color ${t_i}$ to a point in the UV image plane. The RGB values on the UV map, $M$, is determined by the texture color via universal per-pixel alignment according to $C = [c_1, \cdots, c_n]^T \in \mathbb{R}^{n \times 2}$, i.e., $M(c_i)=t_i$. Since this process is reversible, given a UV map, per-vertex texture value can be allocated via a differentiable sampling function $\mathcal{S}$, i.e, $t_i=\mathcal{S}(M, c_i)$, which samples the texture color $t_i$ located in $c_i$, from the UV map, $M$.


\begin{figure*}[t]
\vspace*{-0.9cm}
  \includegraphics[width=\linewidth]{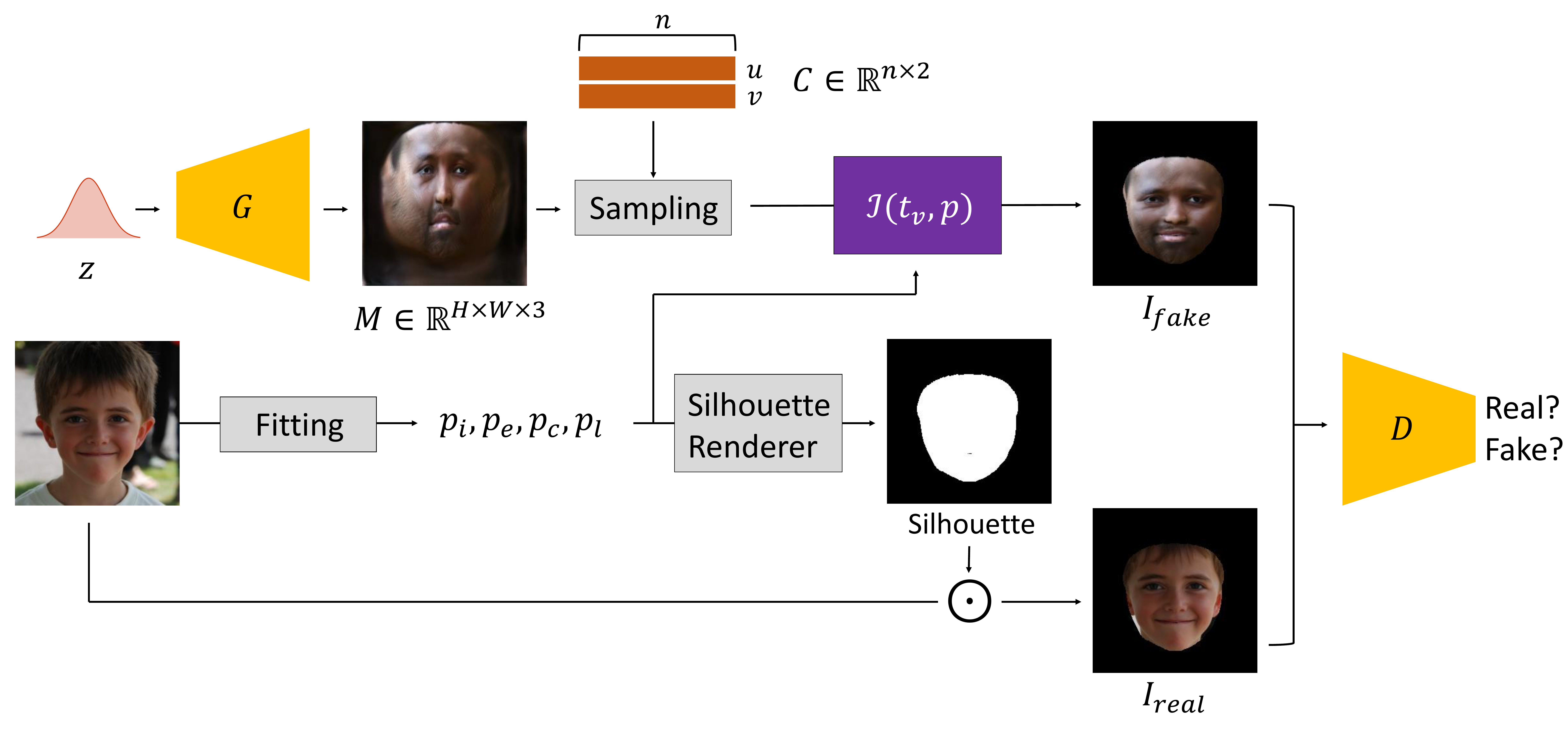}
  \caption{Overview of StyleUV. Given a latent vector, the generator ${G}$ in our framework learns to generate a UV map via the adversarial training with the discriminator $D$. An essential prerequisite for training is obtaining camera and geometry parameters in advance.}
\label{fig:main_fig}
\end{figure*}

\section{Method}
In this section, we elaborate how we design our deep neural networks for the diverse and high-quality UV map generative model. The overview of our framework is introduced in subsection~\ref{subsection:overview}. Subsection~\ref{subsection:uv-generative-model} describes our GANs' framework and a necessity of using a silhouette against our non-trivial problem setting. Subsection~\ref{subsection:3d-reconstruction-networks} and \ref{subsection:image-formation-function} consecutively explain how the 3D representations are obtained from a given image, and how the image formation function works.

\subsection{Overview}\label{subsection:overview}
Our aim is to present a novel generative model for a diverse and high-quality UV map. The most intuitive way to achieve this is to employ a typical GAN framework with real UV maps~\cite{Gecer_2019_CVPR}. However, as written in Section~\ref{sec:introduction}, the diverse and high-quality UV map dataset rarely exists. Hence, we design our framework to learn from real images instead of UV maps. Let our desired generator be a function $G: \mathbb{R}^m \rightarrow \mathbb{R}^{H\times W \times 3}$, which maps a random noise vector $z \in \mathbb{R}^m$ sampled from a Gaussian distribution to a UV map $M \in \mathbb{R}^{H\times W \times 3}$.  
In order to induce the generator ${G}$ to produce a UV map, we leverage the predefined 3DMM parameters and geometric and photometric image formation processes~\cite{liu2019softras,ravi2020accelerating,History} which are differentiable. This fully differentiable model design enables ${G}$ to enjoy a plenty of high-quality training data, i.e., real images. 

Specifically, we adopt one of the off-the-shelf monocular face reconstruction methods using the analysis-by-synthesis strategy to acquire 3DMM parameters for each image, i.e., ${f(I)=p}$, where ${f}$ is a neural network for 3D reconstruction, ${I}$ is a real image and ${p}$\footnote{From now on, $p$ will denote the shape parameter  $p_s$ in (\ref{eq:pca}).} is a 3D shape parameter for the image. We then build a differentiable image formation function $\mathcal{I}(M, C, p)$, taking per vertex RGB values $\{t_i\}_{i=1}^n$, with the parameter ${p}$ as inputs and produces a projected 3D face ${\hat{I}}$ onto the image space. Note that $t_i$ can be obtained from the UV map, $M$, via the grid sampling process, $\mathcal{S}$, i.e., $t_i = \mathcal{S}(M,c_i)$. Lastly, our discriminator ${D}$ is trained via the adversarial loss~\cite{goodfellow2014generative}, which learns to match the distribution of the rendered image to the distribution of a set of the real images.

For the better understanding of our framework and each module, please refer to Fig.~\ref{fig:main_fig} and subsection~\ref{subsection:uv-generative-model}-\ref{subsection:image-formation-function}. 


\begin{figure*}[t]
  \includegraphics[width=\linewidth]{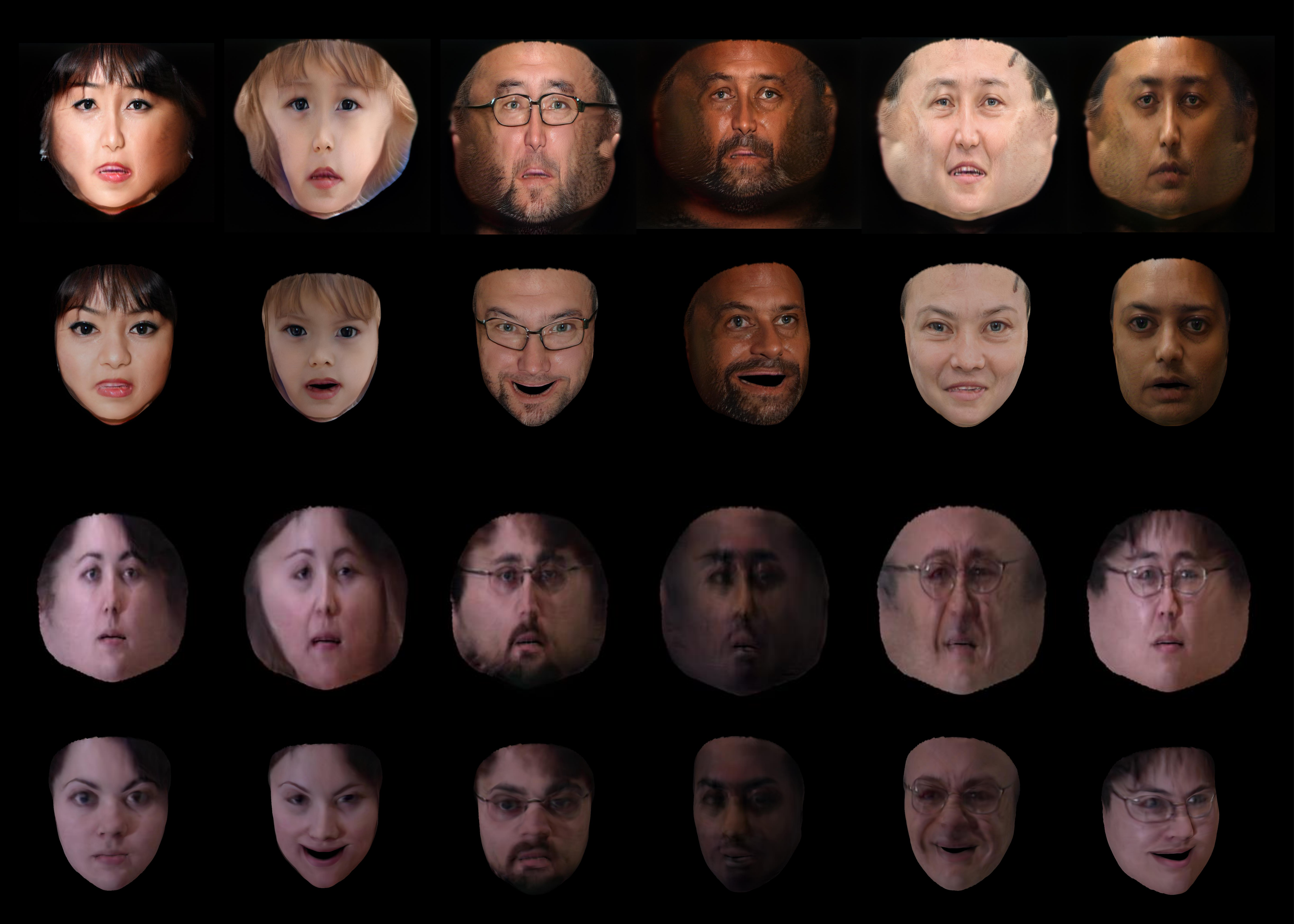}
  \caption{Diverse and high-quality textures generated from our novel texture generative model.}
\label{fig:qualitative_4}
\end{figure*}

\subsection{UV Generative Model}\label{subsection:uv-generative-model}
Numerous GAN-based models have shown its superior performance in generating a high quality image. We choose the state-of-the-art model StyleGAN v2~\cite{Karras_2020_CVPR} 
as our backbone network. Furthermore, from the significantly low FID score it has reported, we can consider the model well covers the diverse real distribution, indicating it stands to reason that we have adopted the model to our framework. 

Generally speaking, a generator ${G}$ takes a latent vector ${z}$ and outputs an image, i.e., ${\hat{I}=G(z)}$. Under the typical adversarial training scheme~\cite{goodfellow2014generative}, taking a real image ${I}$ or a generated image ${\hat{I}}$, a discriminator ${D}$ makes a distribution of the generated images close to that of the real images.
Far from the typical settings of GANs, the discriminator takes a projected image ${\mathcal{I}(M,C,p)}$ in our framework, of which texture $t_i$ is sampled from the UV map, $M$, which comes from the generator $G$, i.e, $M = G(z)$. 


\vspace{1mm} 
\noindent \textbf{Silhouette.} After forwarded into the image formation function ${\mathcal{I}(M,C,p)}$ conditioned on a 3DMM parameter, a generated UV map $M$ forms only the facial region of the projected image ${\hat{I}}$. This means the background information out of the real distribution is entirely unnecessary in the perspective of ${G}$. Possible solutions for this issue are 1) replacing the background region of the projected image ${\hat{I}}$ with a real background and 2) removing the background region of the real image. Although the first one is designed to have the networks ${G,D}$ implicitly focus on the foreground region, we empirically observe that the second one, explicitly designed to match the distribution of only the facial region works better in generating a realistic UV map. We conjecture that the second model design is beneficial in preventing discriminating power of the discriminator from being wasted for discriminating the backgrounds. More detailed comparison will be provided in Section~\ref{section:experiments}.

\subsection{3D Reconstruction}\label{subsection:3d-reconstruction-networks}

Obtaining 3D parameters ${p=(p_i,p_e,p_c,p_l)}$ 
of an image is a crucial part in our framework, where ${p_c,p_l}$ are camera and light parameters, respectively, while $p_i$ and $p_e$ correspond to identity and expression, respectively. In order for this, we adopt widely-used techniques of analysis-by-synthesis to our frameworks, which aims to estimate the parameters that best explain an observed image. To begin with, we prepare facial landmarks~\cite{Kazemi_2014_CVPR} for each image. We then forward an image and its landmarks into one of the off-the-shelf 3D reconstruction networks~\cite{deng2019accurate} to acquire a set of parameters. Lastly, by using the parameter as an initial one, we perform an optimization-based fitting process to have the best parameters for each image. 

\vspace{1mm} 
\noindent \textbf{Energy functions for shape fitting}
Briefly, the fitting process is conducted by minimizing a pixel-level difference and a feature-based difference within the observed image and the projected 3D face. The pixel-level supervision can be formulated as:
\begin{equation}
    E_{pix}(p)=\sum_{(x,y)\in \mathcal{F}}\norm{I_{trg}(x,y)-\mathcal{I}(M,C,p)(x,y)},
\end{equation}
%
%
where ${I_{trg}}$ denotes a target image, ${\mathcal{F}}$ indicates a foreground region (i.e., the face region) and ${\mathcal{I}}$ is an image formation function described in subsection~\ref{subsection:overview}. Note that in this part, the intermediate texture $T = [t_1, \cdots, t_n]^T$, $t_i = \mathcal{S}(M, c_i)$, is the output of the linear texture model ${T_{model}(p_t)}$ in (\ref{eq:pca}). 

On the other hand, the facial-landmark loss is represented as:
\begin{equation}
E_{lm}(p)=\mathbb{E}\norm{t_{trg}-t_{proj}},\; \text{where} \; t_{proj} \subseteq \mathcal{P}(p),
\end{equation}
where $\mathcal{P}(p)$ indicates the 3D to 2D projection using the shape vector $p$ and each $\{t_{trg},t_{proj}\} \in \mathbb{R}^{68\times 2}$ is the 2D landmark coordinates of the target image and a subset of the projected vertices onto the image plane\footnote{The vertex indices of the subset is defined in advance.}.

Combining errors from the above two energy functions, we find the parameters ${p}$ for a given image that best minimizes the loss.

\subsection{Image Formation Function}\label{subsection:image-formation-function}

In order to project a reconstructed 3D mesh to a 2D image, we employ a differentiable renderer~\cite{liu2019softras}, which introduced an approximated gradient for rasterization. A renderer receives scene information including 3D mesh, camera, texture, material and light and outputs an image. Typically, a renderer consists of two main components: rasterizer and shader. First, a rasterizer determines screen-space buffers with triangle IDs and barycentric coordinates for each pixel. A shader computes the color of a pixel by blending the values given by a rasterizer. A light is applied independently to per-pixel with a set of interpolated screen-space buffers. It is modeled by phong shader~\cite{phong} given light position, ambient color and diffuse color in our case. A differentiable renderer enables the integration of rendering into neural networks and makes it possible to propagate losses from 2D images.
 
Specifically, given parameters ${p_i}$ and ${p_e}$, we can estimate the shape by the linear 3DMM model and obtain 3D mesh $S_{model}(p_i, p_e)$. Each vertex in 3D mesh has its texture coordinate assigned in the UV image plane, so we can simply assign the corresponding RGB values from the UV map to each vertex through differentiable sampling function $\mathcal{S}$. The result of the final mesh can be denoted as $M(S_{model},\mathcal{S}({G}(z)))$. It is further projected through the rasterizer given camera parameters $p_c$. Illumination is applied by the Phong shader given $p_l$ and the face normal of the 3D mesh.



%% file: sections/Experiments.tex
\section{Experiments}\label{section:experiments}
In this section, we rigorously verify the superior performance of the proposed StyleUV compared to existing texture models. First, we elaborate implementation details in subsection~\ref{subsection:impl_detail} for a better comprehension of our experimental setup. We then report both quantitative and qualitative comparisons with baselines in subsection~\ref{subsection:quantitative} - \ref{subsection:qualitative-comparisons}. Lastly, we provide analysis on StyleUV justifying our novel frameworks in subsection~\ref{subsection:qualitative-analysis}.



\subsection{Implementation Settings}\label{subsection:impl_detail}

\vspace{1mm} 
\noindent \textbf{Datasets.}
Our primary goal is to train a generator from in-the-wild images for generating high-fidelity and diverse textures. To achieve this goal, we use FFHQ~\cite{Karras_2019_CVPR} as our training dataset, which contains 70,000 high-quality and diverse images at 1,024$^2$ resolution. 
In the testing phase, we perform a qualitative analysis on AFLW2000-3D~\cite{zhu2016face} and in-the-wild images reported in the previous study~\cite{lee2020uncertainty}. We also conduct a quantitative comparison with the existing methods on AFLW2000-3D and CelebA-HQ~\cite{karras2017progressive}.

\vspace{1mm} 
\noindent \textbf{Settings on 3DMM and Fitting Process.}
A combination of the Basel Face Model~\cite{bfm09} and FaceWarehouse~\cite{FaceWarehouse} is used as our 3DMM shape model. With regard to the parameters for 3DMM in our framework, ${p_i\in\mathbb{R}^{80}}$, ${p_e\in\mathbb{R}^{29}}$, ${p_c\in\mathbb{R}^6}$, ${p_l\in\mathbb{R}^6}$ are used as shape, expression, camera and light coefficients respectively.
We remove the neck and ear regions of a face (following~\cite{deng2019accurate}) in order to focus on the facial region only, which utilizes a subset of the 35,709 vertices. The 68 facial landmarks are extracted from the training data by a conventional face-alignment algorithm~\cite{Kazemi_2014_CVPR}. For the fitting process mentioned in subsection~\ref{subsection:3d-reconstruction-networks}, we adopt the Adam optimizer~\cite{kingma2014adam} with a learning rate 0.01. With a good initial point, our fitting process converges in around 3 seconds on an NVIDIA GTX V100 GPU for a single image.
\vspace{1mm} 
\noindent \textbf{GANs' Training Methodologies.}
We employ StyleGAN v2~\cite{Karras_2020_CVPR}, the state-of-the-art GAN framework, to obtain high quality UV maps. 
For accelerating our training process, we adopt the concept of transfer learning~\cite{Wang2018TransferringGG,mo2020freeze} to our framework. Empirically, we observe that this training strategy brings a significant reduction of training time and more stable convergence. Hence, the pre-trained parameters with FFHQ~\cite{Karras_2019_CVPR} images are used to initialize our GAN networks, and the lower layers of the discriminator are frozen.
Hyperparameters and objective functions, including a gradient penalty~\cite{NIPS2017_892c3b1c} and a path length regularization, are the same as in~\cite{Karras_2020_CVPR}.


\subsection{Quantitative Comparisons.}\label{subsection:quantitative}
\noindent \textbf{Experiments using Frechet Inception Distance (FID).}
First of all, we verify the performance of StyleUV from the perspective of generative models. We exploit FID~\cite{heusel2017gans} which is a widely used metric in the GAN literature to evaluate the performance of generative model.
Briefly, it measures the difference between two multivariate Gaussian distributions and is formulated as $\lVert \mu_r-\mu_g\rVert_{2}^{2}+\text{Tr}\left(\Sigma_r+\Sigma_g -2 \left(\Sigma_r\Sigma_g \right)^\frac{1}{2}\right)$, where ${\mu_r,\Sigma_r}$ are the statistics computed from the features of real images and ${\mu_g,\Sigma_g}$ are those from the features of generated images. 
A lower FID score is better because it indicates that the distance between the real and fake distributions is close (in the feature space).

Regarding the experimental settings, since the number of the high-fidelity UV map dataset is limited, we measure FID at the image level.
To focus on measuring the quality of the foreground regions, we mainly measure FID scores with foreground-masked images.
Concretely, we use CelebA-HQ~\cite{karras2017progressive} dataset for a fair comparison with the baseline; StyleUV learns from FFHQ dataset while GANFIT learns from the UV dataset. We first randomly sample 30,000 images and corresponding 3D parameters, i.e., ${(p_i,p_e,p_c,p_l)}$ out of the entire dataset.
We then acquire two statistics of the real images, one for the real image itself and the other for the foreground-masked real image. Next, we obtain two sets of 30,000 generated images by forwarding randomly sampled latent vectors into StyleUV and GANFIT, respectively. After projecting the generated images onto the image plane with corresponding parameters, we acquire the statistics from the projected images which covers only the foreground regions. Lastly, we compute four FID scores out of a combination of features from the generated images and the real images, i.e., ours \emph{v.s.} real, ours \emph{v.s.} masked real, etc.

With regard to GANFIT~\cite{Gecer_2019_CVPR} (a baseline model based on GANs), we reproduce the paper by fine-tuning a FFHQ-pretrained StyleGAN v2 model with the publicly available UV map dataset~\cite{deng2018uv}. Note that the GAN-based works can be compared in this experiment.


\begin{table}[]
\caption{Evaluations on FID score. C-HQ denotes CelebA-HQ dataset~\cite{karras2017progressive}. FID is computed from a comparison between distributions of the real images and the projected images without background. Masked FID is a distance between the distributions of the foreground-masked real images and the projected images. The spatial resolution of the images used in this experiment is ${256^2}$.}
\begin{center}
\vspace{-5mm}
\begin{tabular}{l|llll}
\hline
Measure & \multicolumn{2}{c}{FID} & \multicolumn{2}{c}{Masked FID} \\ \hline
Dataset & FFHQ       & C-HQ       & FFHQ           & C-HQ          \\ \hline
GANFIT~\cite{Gecer_2019_CVPR}       &    226.9        &     214.4       &     49.9            &     68.3           \\
StyleUV(\emph{w.o.} mask)       &     198.7       &      201.8      &       14.3         &     51.9          \\
StyleUV       &     166.8       &     163.2       &        11.3        &       34.0        \\ \hline
\end{tabular}
\label{Tab:experiments-fid-score}
\end{center}
\vspace{-4mm}
\end{table}


As shown in Table~\ref{Tab:experiments-fid-score}, our method outperforms the baseline method in terms of both FID and Masked FID. This comparison verifies that the generated UV maps from StyleUV are high-quality and cover the diverse nature of real faces. Moreover, superior performance of StyleUV to that trained without masks justifies the importance of the foreground mask.

\vspace{1mm} 
\noindent \textbf{Experiments on 3D Reconstruction.}
For exploring a possible use case of StyleUV, we further quantitatively evaluate the performance of our model with respect to the texture quality by measuring the reconstruction errors. In order to solely measure the performance of the texture model, we conduct experiments on the ALFW2000-3D dataset, which contains ground-truth geometry and camera information. Given a facial image with ground-truth geometry and camera coefficients, we optimize the texture and light parameters by minimizing the $L_1$ distance between the rendered image and the corresponding real image. Once the parameters are optimized for a given image, we measure the average $L_{2,1}$ reconstruction error over the facial region of an image. 

As shown in Table~\ref{Tab:experiments-L2,1}, StyleUV achieves the lowest $L_{2,1}$ error by a large margin out of the prominent texture models.
The impressive performance of StyleUV in 3D reconstruction verifies that it has a good texture representational power,  implying that textures from StyleUV are highly reliable. 
Furthermore, it is worth noting that the GANFIT model, which has the same network architecture as ours, performs worse than our model because it learns from a limited number of UV maps while StyleUV learns from a large number of image datasets in our novel framework.


\begin{table}[tbp]
\begin{center}
\caption{Quantitative results on ALFW2000-3D. The average $L_{2,1}$ reconstruction errors are computed over the facial portion of each image.}
\vspace{-2mm}
\begin{tabular}{lclll}
\hline
Method                         & \multicolumn{4}{c}{Reconstruction error ($L_{2,1}$)} \\ \hline
Linear~\cite{zhu2016face}                 & \multicolumn{4}{c}{0.1287}                      \\ 
Nonlinear~\cite{on-learning-3d-face-morphable-model-from-in-the-wild-images}              & \multicolumn{4}{c}{0.0427}                      \\ 
Nonlinear + GL + Proxy~\cite{tran2019towards} & \multicolumn{4}{c}{0.0363}                      \\ 
GANFIT~\cite{Gecer_2019_CVPR}                 & \multicolumn{4}{c}{0.0122}                      \\ 
Lee et al.~\cite{lee2020uncertainty}             & \multicolumn{4}{c}{0.0317}                      \\ 
StyleUV(ours)                  & \multicolumn{4}{c}{\textbf{0.0106}}              \\ \hline
\end{tabular}
\label{Tab:experiments-L2,1}
\end{center}
\vspace{-2mm}
\end{table}

To analyze the performance of StyleUV from diverse angles, we also evaluate a quality of the reconstructed images at the perceptual level, i.e., we compute the cosine similarity between the feature vectors of the reconstructed image and its corresponding real image.
FaceNet~\cite{schroff2015facenet} pretrained on CASIA-Webface~\cite{yi2014learning} is used as our feature extractor.
As seen in Table~\ref{Tab:experiments-Cosine-Similarity}, the cosine similarity of StyleUV is higher than other baselines. This indicates that the reconstructed 3D face using our texture model is perceptually similar to the original image.


\begin{table}[tbp]
\begin{center}
\caption{Quantitative results on ALFW2000-3D. The average $L_{2,1}$ reconstruction errors are computed over the facial portion of each image.}
\begin{tabular}{lclll}
\hline
Method                         & \multicolumn{4}{c}{Cosine similarity} \\ \hline
Linear~\cite{zhu2016face}                 & \multicolumn{4}{c}{0.7032}                      \\ 

GANFIT~\cite{Gecer_2019_CVPR}                 & \multicolumn{4}{c}{0.7365}                      \\ 
StyleUV(ours)                  & \multicolumn{4}{c}{\textbf{0.7494}}              \\ \hline
\end{tabular}
\label{Tab:experiments-Cosine-Similarity}
\end{center}
\vspace{-2mm}
\end{table}


    

\subsection{Qualitative Comparisons}\label{subsection:qualitative-comparisons}

\noindent \textbf{UV Map Generation.}
To begin with, we compare the quality of UV maps between the generative models. As demonstrated in Fig.~\ref{fig:qualitative_4}, UV maps from StyleUV show superior performance over those from GANFIT. The first and the second rows are the UV map and its projecte image from StyleUV and the third and the fourth rows are those from GANFIT. It is readily noticed that the diversity and the texture quality of StyleUV is much better than GANFIT. The UV maps from StyleUV cover diverse races and ages while those from GANFIT show the limited diversity including monotonous complexions and the narrow scope of age. 

It is worth noting that although StyleUV and GANFIT are built from the same model architecture, UV maps from ours clearly show more diverse and high-quality UV maps. We believe these enhancements come from our novel framework. That is, GANFIT only learns from the UV dataset limited in terms of both quality and quantity while StyleUV is trained with a large-scale, high-quality image dataset.

\begin{figure}[h]
  \includegraphics[width=\linewidth]{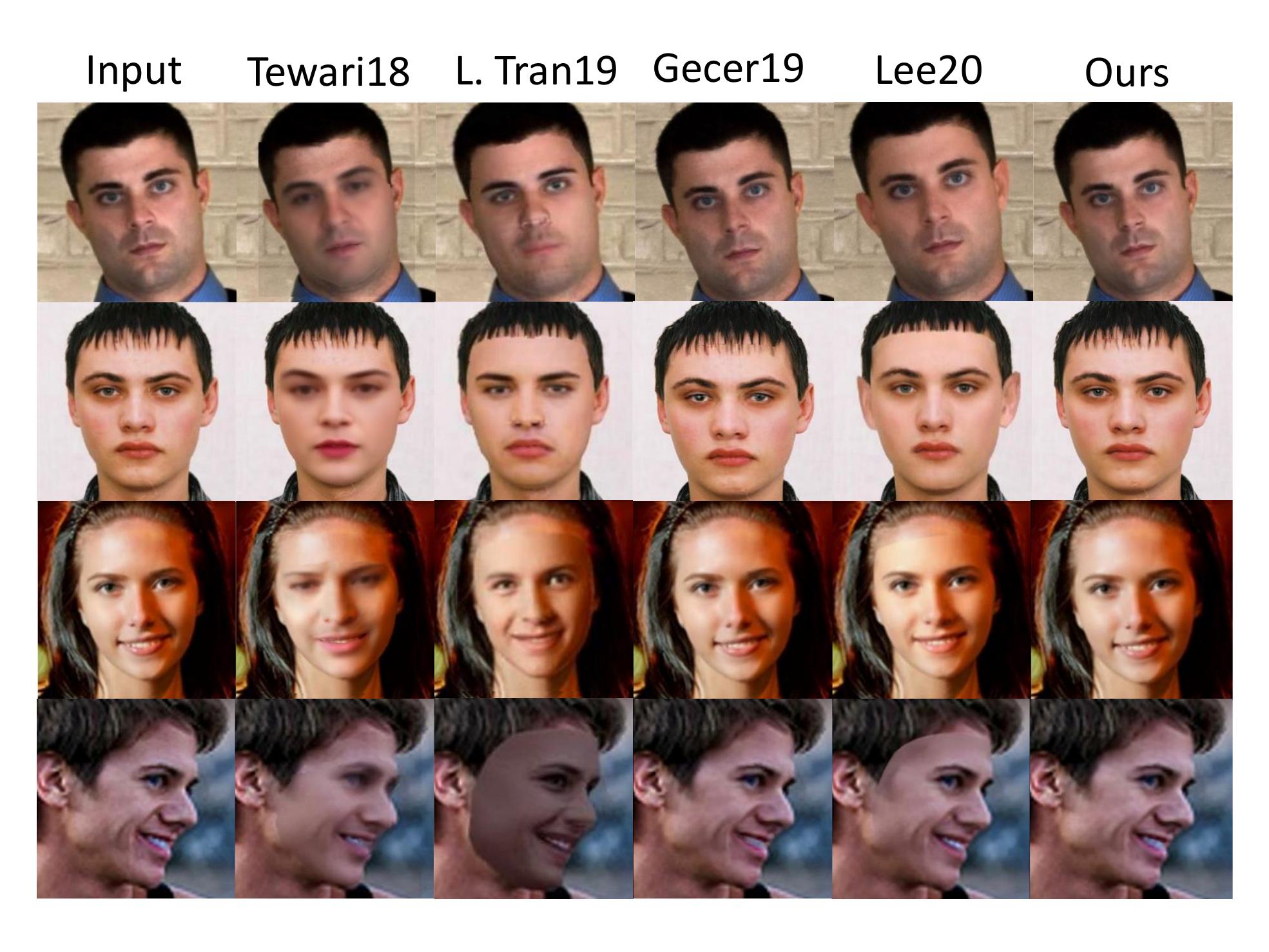}
  \caption{Comparisons on reconstruction capability with baselines on ALFW2000-3D dataset.}
\label{fig:figure_1}
\end{figure}

\vspace{1mm} 
\noindent \textbf{Reconstruction.}
We further present qualitative results of the experiment on face reconstruction to visually verify the superior performance of StyleUV. Figure~\ref{fig:figure_1} shows the superior and reliable aspect of GANs as a texture model. The first column indicates the input image and the others represent the reconstructed images through different texture models. As seen in the first and the third columns from right, the strong representational power of GANs leads to outstanding fitting performance. Other texture models wrongly estimate inappropriate textures for given images, yielding conspicuous color differences with the projected 3D face and the other part of the face in the image.


\subsection{Analysis on StyleUV}\label{subsection:qualitative-analysis}

\begin{figure}[t]
  \includegraphics[width=\linewidth]{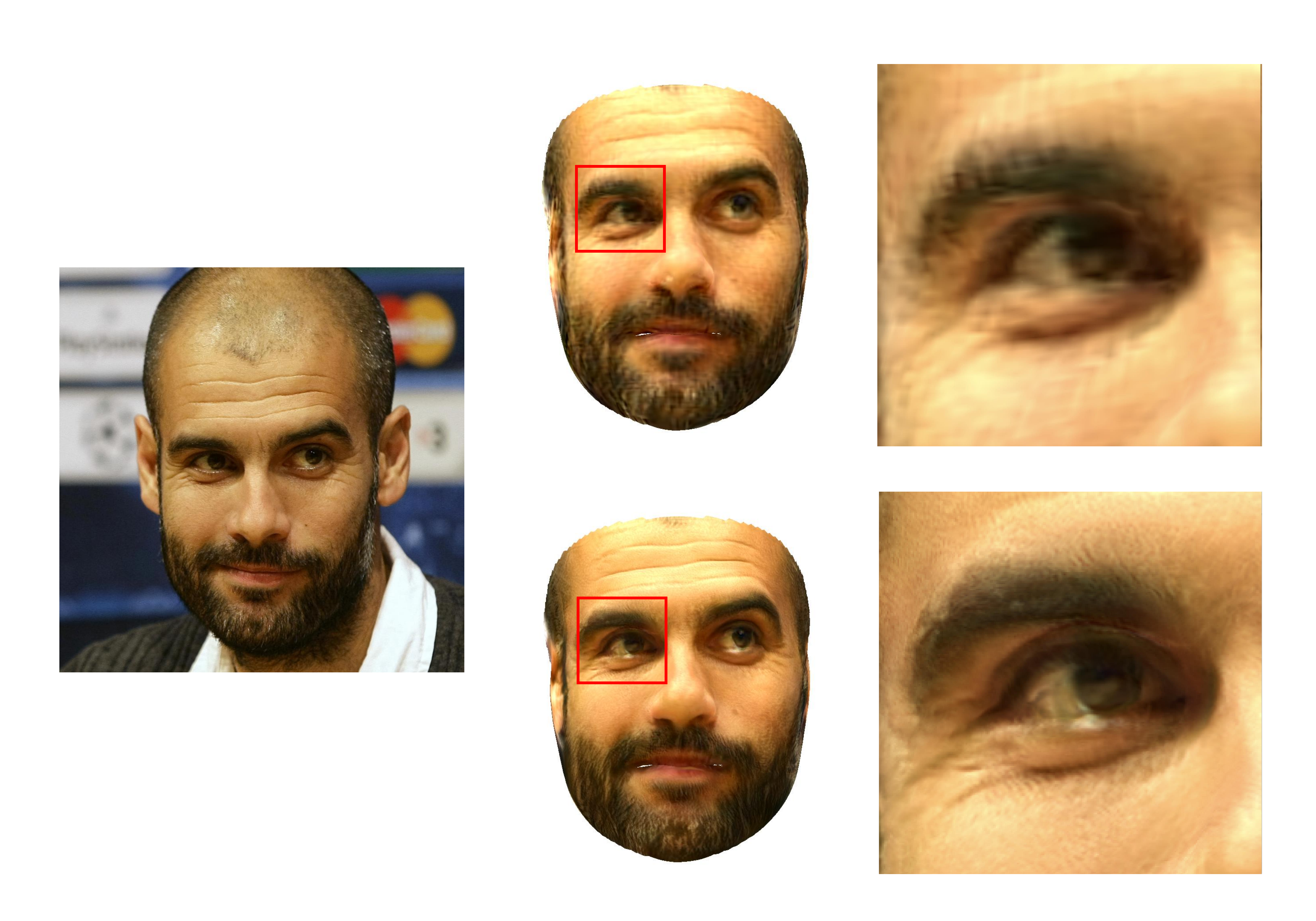}
  \caption{Benefits of high-quality texture model. Top: GANFIT (learned from ${256^2}$), Bottom: StyleUV (learned from ${512^2}$).}
\label{fig:figure_3}
\end{figure}

\noindent \textbf{High-quality UV Generation.}
One of the strong benefits of StyleUV over GANFIT is that our model can learn to generate the high-fidelity UV maps without any ground-truth UV maps. To show a practical example of the strength of a high-quality texture model, we compare the reconstruction quality between StyleUV trained with images of ${1024^2}$ and GANFIT trained with UV maps of ${256^2}$. As shown in Fig.~\ref{fig:figure_3}, StyleUV generates clear and vivid textures while the textures from GANFIT contain a few of artifacts. This experiment implies that our proposed framework, which enables GANs to learn from images has a practical benefit.


\begin{figure}[t]
  \includegraphics[width=\linewidth]{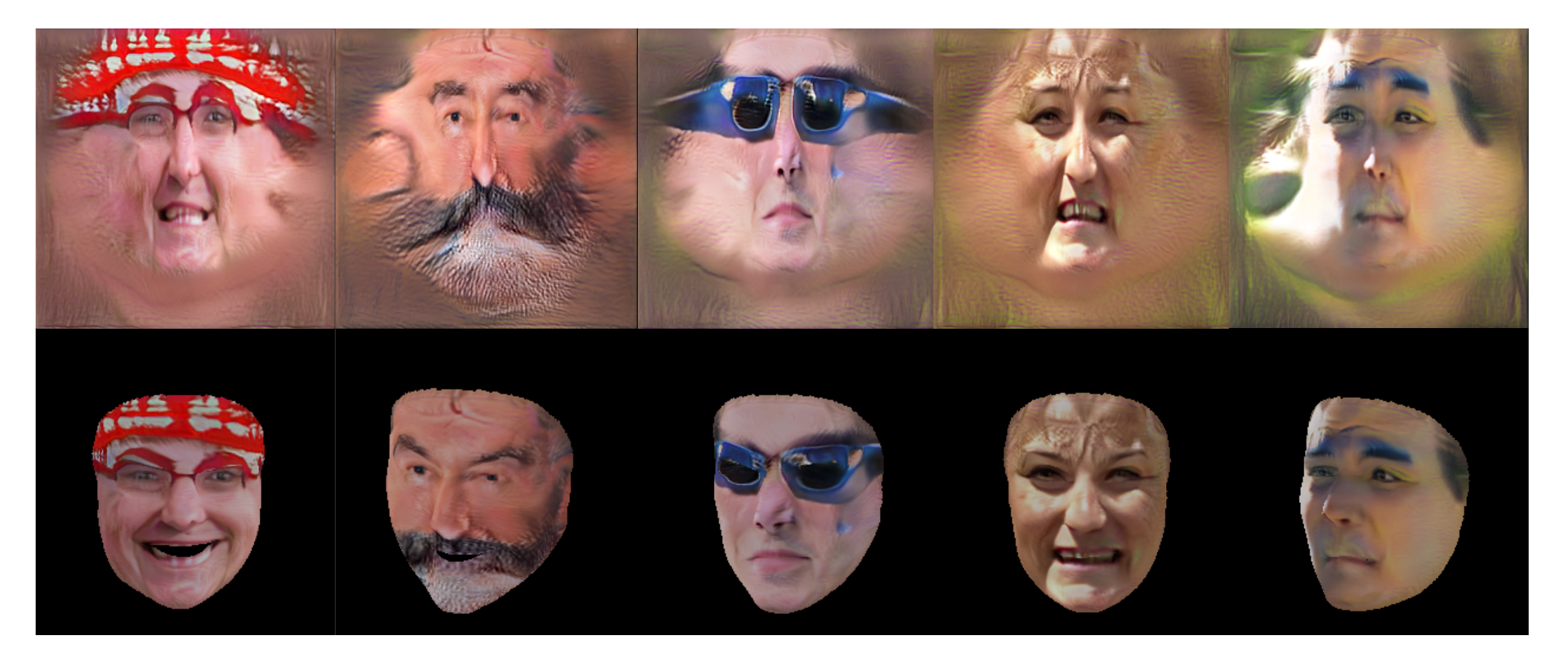}
  \caption{Visualizion of generated UV maps from a StyleUV trained without the silhouette mask. The first row is the generated UV maps and the second row indicates rendered images.}
\label{fig:qualitative_5}
\end{figure}

\vspace{1mm} 
\noindent \textbf{Necessity of Silhouette.}
We additionally conduct a qualitative analysis on the foreground mask. As written in subsection~\ref{subsection:uv-generative-model}, the foreground mask plays an important role in our framework to properly train GANs. Fig.~\ref{fig:qualitative_5} represents the UV map results of StyleUV \emph{w.o.} mask. The results show that a StyleUV learned without the foreground mask is degraded because of the limited capability of drawing a realistic facial UV map with fine details. For example, the hat in the first column and the sunglasses in the third column are unrealistically generated. Moreover, other details including wrinkles and clear facial features, e.g., eyes, mouth, etc., are obviously degraded. Throughout the observations, including the quantitative comparisons in Table~\ref{Tab:experiments-fid-score}, we justify the necessity of silhouette.

%% file: sections/Conclusion.tex

\section{Conclusion}
In this paper, we propose a novel generative model that is able to generate photo-realistic UV maps that span the diversity of real faces.
We present a non-trivial GAN-based training framework that includes a novel rendering-based adversarial training approach with the foreground silhouette, which enables StyleUV to generate photo-realistic and diverse UV maps.
The superior performance of StyleUV over the prominent baselines is demonstrated by rigorously designed experiments. Regarding limitations, the performance of our frameworks highly relies on a predefined geometry and a light condition. We suppose that if those predefined components can be obtained or refined during the training in an end-to-end manner, the performance of StyleUV would be enhanced. Another possible research direction is to leverage the symmetric information of the UV map to improve the performance of StyleUV. We hope StyleUV will be a useful stepping stone for the future research on texture generative model.